\documentclass[conference]{IEEEtran}
\ifCLASSINFOpdf
\else
\fi
\usepackage{graphicx}
\usepackage{cite}
\usepackage{algorithm}
\usepackage[noend]{algpseudocode}
\usepackage{algorithmicx}
\usepackage[draft]{hyperref}
\usepackage{subcaption}

\usepackage{amsmath}
\usepackage{amssymb}
\begin{document}
%
\title{Machine Learning for Pre/Post Flight UAV Rotor Defect Detection Using Vibration Analysis
}
%
%
%

\author{Alexandre~Gemayel, Dimitrios~Michael~Manias, and~Abdallah~Shami \\ The Department of Electrical and Computer Engineering, Western University\\ \{agemayel, dmanias3, Abdallah.Shami\}@uwo.ca}

%
%

\markboth{IEEE GLOBECOM 2024}%
{Gemayel \MakeLowercase{\textit{et al.}}:Machine Learning for Pre/Post Flight UAV Rotor Defect Detection Using Vibration Analysis}
%



\maketitle

\begin{abstract}
Unmanned Aerial Vehicles (UAVs) will be critical infrastructural components of future smart cities. In order to operate efficiently, UAV reliability must be ensured by constant monitoring for faults and failures. To this end, the work presented in this paper leverages signal processing and Machine Learning (ML) methods to analyze the data of a comprehensive vibrational analysis to determine the presence of rotor blade defects during pre and post-flight operation. With the help of dimensionality reduction techniques, the Random Forest algorithm exhibited the best performance and detected defective rotor blades perfectly. Additionally, a comprehensive analysis of the impact of various feature subsets is presented to gain insight into the factors affecting the model's classification decision process.
\end{abstract}


%
\IEEEpeerreviewmaketitle

\section{Introduction}
%
%
%
%
\IEEEPARstart{S}{mart} Cities represent the pinnacle of modern innovation and describe a data-driven ecosystem that is adaptive and responsive to changes in its environment. Smart Cities lie at the crossroads of various enabling technologies, such as the Internet of Things (IoT), Machine Learning (ML) and Artificial Intelligence (AI) for automation, as well as Smart Sectors, including transportation, energy and infrastructure \cite{IBM}. Smart Cities are characterized by constant monitoring and vast volumes of data used to guide decision-making processes to improve the quality of life for all residents. One critical distinction is that a resident of a Smart City is not limited to human constituents and encompasses the natural environment within its boundaries. \par

The Unmanned Aerial Vehicle (UAV), commonly known as the drone, is one of the exciting aspects of future Smart Cities. UAVs equipped with sensors and cameras are vital data collection entities that will relay critical information that will ultimately help inform decision-making agents of the system's current state. Some anticipated applications of UAVs in Smart Cities include infrastructure inspection, traffic monitoring, environment monitoring, as well as delivery and taxi services \cite{mohamed2020unmanned}. Given the role of UAVs in Smart Cities, their reliability is a critical priority. The topic of UAV reliability considers multiple aspects ranging from communication to component reliability. Another important reliability aspect relates to security as UAVs are prone to cyberattacks to steal information or gain control \cite{mohsan2023unmanned}. In order to ensure reliability, the hardware, software, and communication of the UAV must constantly be monitored for performance changes and deviation from expected behavior. \par

The work presented in this paper deals with UAV health monitoring from the hardware perspective as it explores the use of signal processing and ML techniques to identify rotor blade defects through a vibrational analysis. Using two mounted sensors, vibrational data from the UAV is constantly relayed and processed. The processing of this data consists of the extraction of time and frequency domain features that are used to train ML models to classify the performance as normal or defective. This work specifically considers the pre- and post-flight operation of the UAV and works towards fully automating the rotor condition monitoring process. \par

The remainder of the paper is structured as follows. Section II outlines related work in the field. Section III presents the system model. Section IV discusses the methodology. Section V presents and analyzes the results. Finally, Section VI concludes the paper and outlines avenues for future work.

\section{Related Work}
The following outlines work done in the field of UAV vibrational analysis and rotor defect detection. Simsiriwong and Sullivan \cite{simsiriwong2012experimental} conduct a vibrational analysis on a UAV wing made from composite material. The authors consider frequency-based features by leveraging frequency response functions, specifically the Discrete Fourier Transform (DFT), focusing on acceleration. The experiments leveraged a shaker table and induced oscillations.
Bektash and Cour-Harbo \cite{bektash2020vibration} leverage a UAV vibration analysis for anomaly detection. Their objective is to use the analysis results to gain insight into the health of the drone from a mechanical perspective. The authors explored various payloads and hovering altitudes. The authors suggest an acceptable performance region determined by statistical thresholding for normal operation.
Al-Haddad \textit{et al.} \cite{al2023investigation} consider using a vibrational analysis for UAV rotor blade fault detection. They posit that defective rotors affect flight balance and use this to motivate their work. They leveraged the Fast Fourier Transform (FFT) algorithm to convert from the time to the frequency domain. The authors visually explore the differences between normal and damaged rotor flights by plotting the frequency spectrum.
Banerjee \textit{et al.} \cite{banerjee2020flight} explore UAV in-flight anomaly detection of vibration signals. The authors discuss various motor-specific failure modes along with their causes and criticalities. The authors use the FFT algorithm to convert the time-domain signal to the frequency domain and then generate Power Spectral Density (PSD) plots, which are then used to develop metrics that are compared to predetermined thresholds to indicate the presence of an anomaly.
Bondyra \textit{et al.} \cite{bondyra2017fault} leverage signal processing methods and the Support Vector Machine (SVM) to identify rotor faults caused by the propulsion system. They extract features from the time-domain signal using methods, including the FFT, to train the SVM classifier. The authors evaluate various defect types and flight patterns. \par
While many of these works explore various aspects of rotor defect detection through various conditions, several gaps in the literature must be addressed. Firstly, many analyses apply to the listed conditions and are not generalizable to other flight conditions. Secondly, there is a lack of high-quality, extensively documented, publicly available data that can be used by the research community to further research in the field. Finally, to the best of our knowledge, a comprehensive analysis of the performance of various ML algorithms and the features contributing to their performance does not currently exist. To this end, the contributions of this work are as follows:
\begin{itemize}
\item A signal processing-based vibrational analysis during pre and post-flight operation that is used to automate rotor fault detection across various defect types through ML.
\item The release of publicly available vibrational data to the research community for democratizing work in this field.
\item A comprehensive analysis of the effect of various time and frequency-domain features on ML model performance.
\end{itemize}

\section{System Model}
The following section outlines the system model built in this work. The Helipal Storm 4 is the drone used. It is a Fully Assembled Storm Drone 4 Flying Platform (CC3D, Radio Link of 2.4\textit{GHz} AT9, Radio System w/ R9D 9-Ch Receiver, 11.1\textit{V} 2200\textit{mAh} 35C Li-Po Battery. The drone is positioned in a stable state, not flying but is turned on in its location. The blade, under investigation, is positioned on the upper-right wing section of the UAV. \par
\begin{figure} [!hbtp]

\begin{subfigure} {0.9\columnwidth}
\hspace{0.025\columnwidth}    \centerline{\includegraphics[width=\columnwidth]{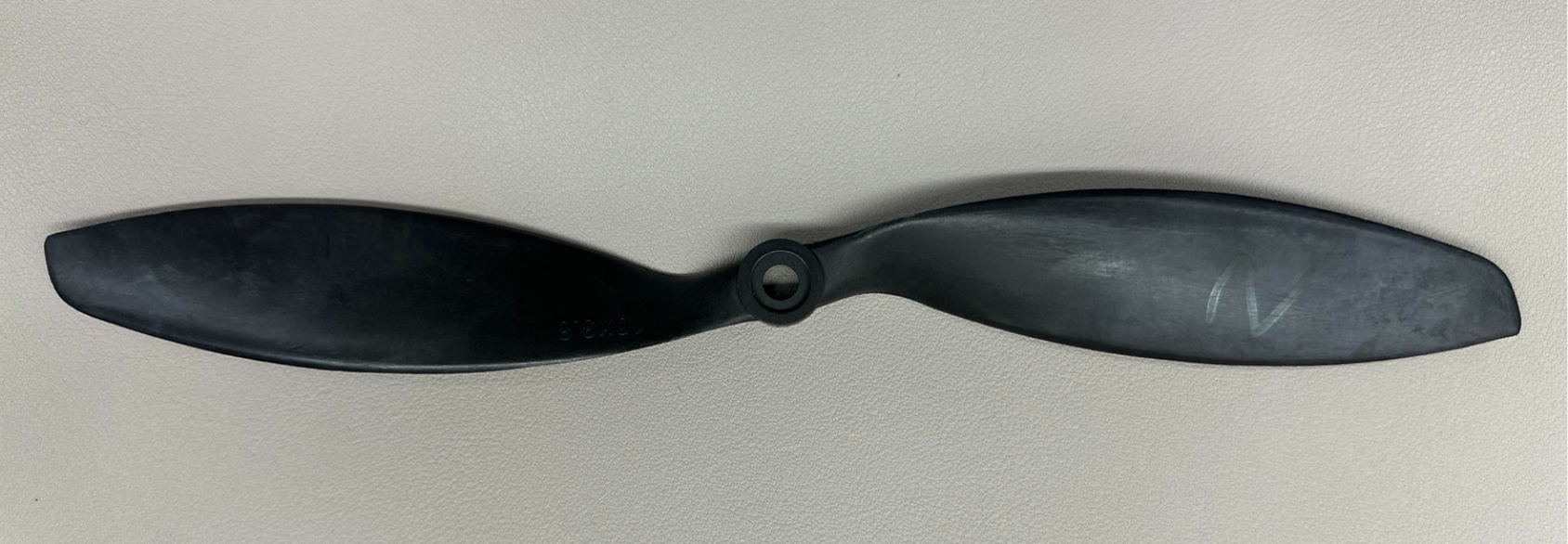}}
    \caption{Normal Rotor}
    \label{normal}
\end{subfigure}

\begin{subfigure} {0.9\columnwidth}
    \hspace{0.025\columnwidth} \centerline{\includegraphics[width=\columnwidth]{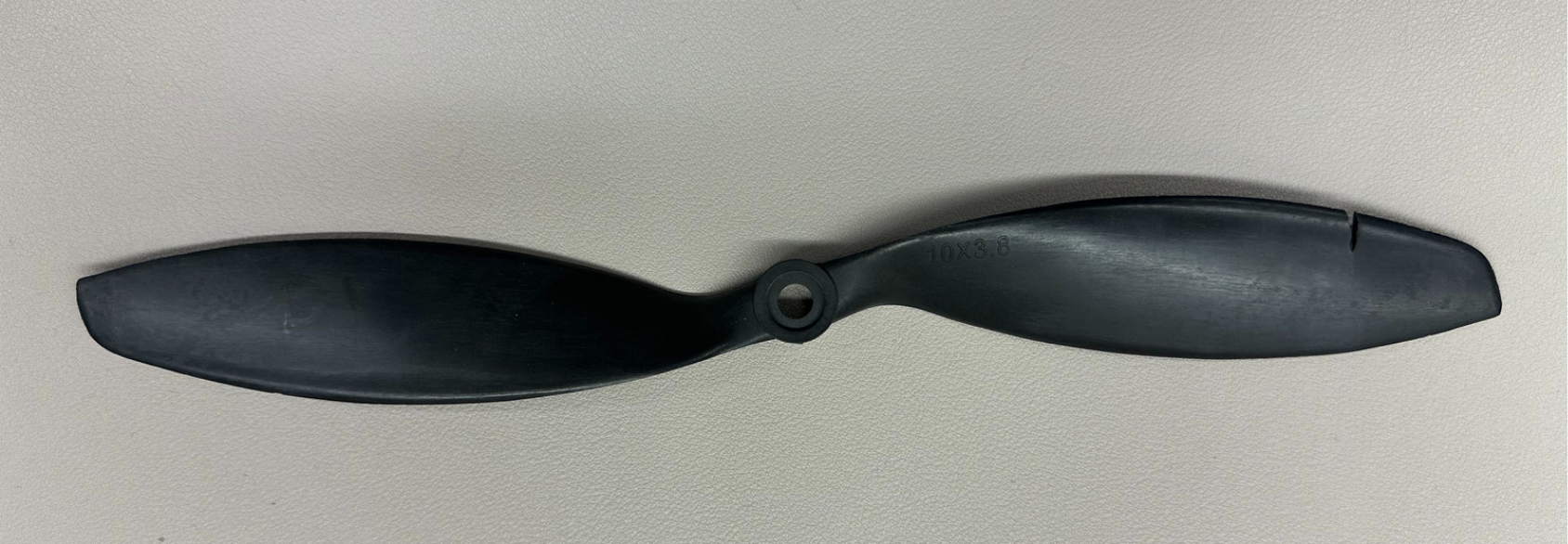}}
    \caption{Rotor Defect Type 1}
    \label{D1}
\end{subfigure}

\begin{subfigure} {0.9\columnwidth}
    \hspace{0.025\columnwidth} \centerline{\includegraphics[width=\columnwidth]{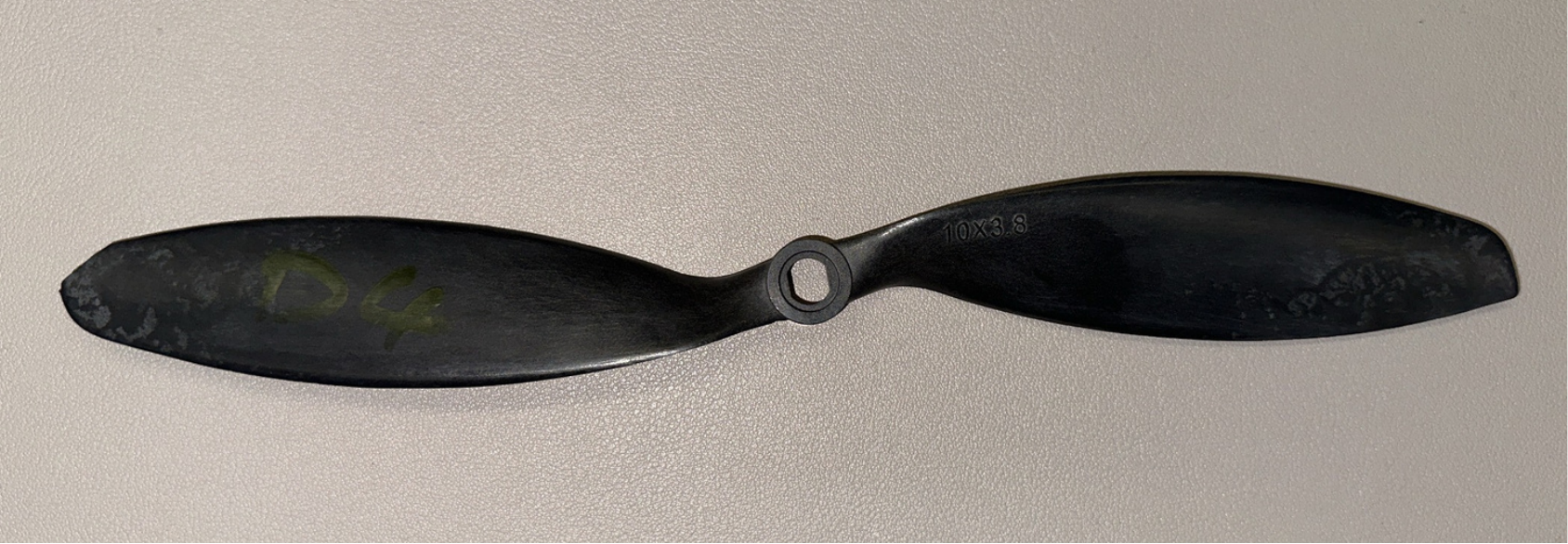}}
    \caption{Rotor Defect Type 2}
    \label{D2}
\end{subfigure}

\begin{subfigure} {0.9\columnwidth}
    \hspace{0.025\columnwidth} \centerline{\includegraphics[width=\columnwidth]{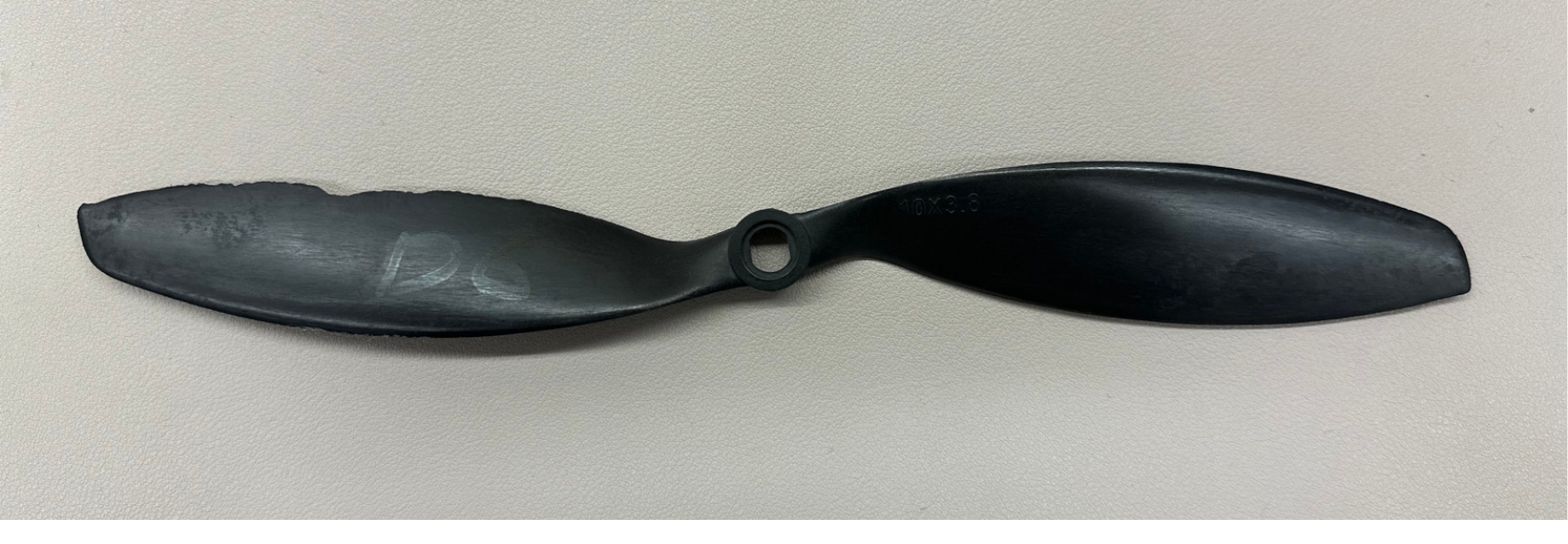}}
    \caption{Rotor Defect Type 3}
    \label{D3}
\end{subfigure}

\caption{Normal \textit{vs.} defective rotor types.}
\label{Rotors}

\end{figure}

 
In this work, data collection experiments are conducted using four types of blades depicted in Fig. \ref{Rotors}. The first blade type exhibits the base condition, without any defect, and is considered the normal state of our system as seen in Fig. \ref{normal}. The second type of blade is defective with a crack introduced at one end, as seen in Fig. \ref{D1}. The third blade type consists of a defect where one end has been trimmed as seen in Fig. \ref{D2}. The fourth blade type is defective with scratches and misshaped on one side, as seen in Fig. \ref{D3}. For the purposes of these experiments, two variations of each blade defect type are used to ensure variety in the data. \par

Two ADXL345 sensors are used to record the vibrations. One of them is positioned at the center of the drone, and the second one is positioned at the top-right wing of the drone. The ADXL345 is a compact, thin, ultralow power, three-axis accelerometer that can measure up to ±8\textit{g} with a high resolution (13-bit). ADXL345 monitors the static acceleration of gravity as well as the dynamic acceleration brought on by motion or shock. Its 3.9 \textit{mg/LSB} high resolution makes it possible to measure inclination variations of less than 1.0°. \par
The System on Chip (SoC) microcontroller ESP32, with dual-mode Bluetooth 4.2, built-in Wi-Fi 802.11 b/g/n, and a range of peripherals is used. It is a more advanced version of the 8266 chips, with two cores that can be clocked at different frequencies up to 240 \textit{MHz}. The ESP32 is powered by a 40 \textit{nm} processed Xtensa LX6 CPU, and has two cores, each with independent control. 520 \textit{kB} of on-chip SRAM is available for data and instructions. 
The ESP32 microcontroller is connected to the local machine using a serial connection. For data collection, PuTTY, an open-source terminal emulator and serial console, is used. Both the Central and the Outer sensors have a sampling rate of 800 \textit{Hz} and a range of ±8\textit{g}. When the data collection process is initiated, every vibration event is captured along with its unique timestamp, sensor identifier, and the associated x, y, and z-axis signal values.

\section{Methodology}
The following section outlines the methodology followed
for this work, including the data collection and dataset creation process, feature extraction, as well as machine learning implementation.
\subsection{Data Collection}
Regarding the data collection process, experiments are conducted by turning the drone on and capturing its vibrations, as described in the previous section, using the Central and Outer mounted sensors. Every experiment is conducted for 1 minute. After the experiment duration is complete, the recorded vibrations, during this period, are exported and stored as a data file. This data file is then processed into a .csv formatted file, where every vibration recorded corresponds to a row. The records in this file consist of data, including what sensor (Central or Outer) collected the data, the timestamp, and the associated X, Y, and Z signal values. 8 experiments were conducted on the Normal blade, serving as the Normal UAV Condition, and 3 experiments on each of 6 Defective blades (2 of each defect type).\footnote{data will be made available on \url{https://github.com/Western-OC2-Lab/} pending the acceptance of this paper.} 

\subsection {Feature Extraction}
After all experiments were conducted, each data file is processed using a feature extraction pipeline. As a first step, the experiment file is uploaded, and the sensor (Central or Outer) is selected. In every experiment, vibrations are grouped into subsets of 800 vibrations, corresponding to 1 second of data given the UAV sampling frequency of 800\textit{Hz}. The grouping of data in 1 second intervals was selected to align with the sampling frequency and minimize reporting delays; future work will explore varying this period. The following outlines the various features extracted.  

The magnitude of every vibration record, $r$ is calculated using Eq. \ref{magnitude}, where $X_i$, $Y_i$, and $Z_i$ denote the vibration along the $x$, $y$, and $z$ axes, respectively.

\begin{equation}
    |r| = \sqrt{X_{i}^{2} + Y_{i}^{2} + Z_{i}^{2}}
    \label{magnitude}
\end{equation}

In the time domain, features such as the Amplitude, Mean, and Standard Deviation of $X$, $Y$, and $Z$ are extracted. Equation \ref{amplitude} is used to calculate the amplitude, $A$, where $max$ and $min$ refer to the vibration signal's maximum and minimum values during the observation period. Equation \ref{mean} is used to calculate the mean of the vibration signal during the observation period, denoted by $\bar{x}$, where $x_i$ denotes a sample value and $n$ denotes the number of samples. Finally, Eq. \ref{standard_dev} is used to calculate the standard deviation, $\sigma$, where $x_i$ denotes a sample value, $\bar{x}$ denotes the mean, and $n$ denotes the number of samples.

\begin{equation}
    A = \frac{max - min}{2}
    \label{amplitude}
\end{equation}

\begin{equation}
    \bar{x} = \frac{\sum\limits_{i=1}^{n}{x_i}}{n}
    \label{mean}
\end{equation}

\begin{equation}
    \sigma = \sqrt{\frac{\sum\limits_{i=1}^{n}{(x_i - \bar{x})^2}}{n}}
    \label{standard_dev}
\end{equation}

In the frequency domain, all Short Time Fourier Transform (STFT) features are extracted for $X$, $Y$, and $Z$ using Eq. \ref{stft}, where $x(n)$ is the input signal at time $n$, $w(n)$ is the window function, $X_{m}(\omega)$ denotes the Discrete Time Fourier Transform (DTFT) of windowed data centered about time $mR$, and $R$ denotes the sample hop size between successive DTFTs.

\begin{equation}
    X_{m}(\omega) = \sum\limits_{n=-\infty}^{\infty}x(n)w(n-mR)e^{-j{\omega}n}
    \label{stft}
\end{equation}

The Shannon Entropy $H(x)$ is calculated using Eq. \ref{entropy}, where $P(x_i)$ denotes the probability of a single event.

\begin{equation}
    H(x) = -\sum\limits_{i=1}^{n}P(x_i)\log_{b}P(x_i)
    \label{entropy}
\end{equation}

The Wavelet Packet Transform is also used to create a set of features. This transform improves upon wavelet-based decomposition by capturing both high and low frequency components \cite{aburakhia2024intersection}. 

The Spectral Centroid, $\mu_1$, is calculated using Eq. \ref{centroid}, where $f_k$ is the frequency corresponding to bin $k$, $s_k$ is the spectral value at bin $k$, and $b1$ and $b2$ are the band edges across which the centroid is calculated.

\begin{equation}
    \mu_1 = \frac{\sum\limits_{k=b_1}^{b2}f_{k}s_{k}}{\sum\limits_{k=b_1}^{b_2}s_k}
    \label{centroid}
\end{equation}

Spectral Skewness $\mu_3$ is the third-order moment and measures spectrum asymmetry around the centroid as expressed through Eq. \ref{skew}. The same base notations apply as in Eq. \ref{centroid} with the addition of $\mu_1$ representing the spectral centroid and $\mu_2$ representing the spectral spread.

\begin{equation}
    \mu_3 = \frac{\sum\limits_{k=b_1}^{b2}(f_{k}-\mu_1)^{3}s_{k}}{(\mu_{2})^3\sum\limits_{k=b_1}^{b_2}s_k}
    \label{skew}
\end{equation}

Table \ref{features} quantifies the number of each type of extracted feature on a per-axis, per-sensor, and per-record basis.

\begin{table}[!hbtp]
\caption{Number of features extracted based on feature types}
\label{features}
\begin{tabular}{|c|c|c|c|}
\hline
\textbf{Type}               & \textbf{Per Axis} & \textbf{Per Sensor} & \textbf{Per Record} \\ \hline
\textbf{Time-Domain}        & 4                 & 12                  & 24                  \\ \hline
\textbf{STFT}               & 4,617             & 13,851              & 27,702              \\ \hline
\textbf{Wavelet}            & 8                 & 24                  & 48                  \\ \hline
\textbf{Spectral Centroid}  & 1                 & 3                   & 6                   \\ \hline
\textbf{Frequency Skewness} & 1                 & 3                   & 6                   \\ \hline
\textbf{Total}              & 4,631             & 13,893              & 27,786              \\ \hline
\end{tabular}
\end{table}

\subsection{Machine Learning}

Once all the features are extracted, the ML process begins. The first stage in this process involves splitting the data into training and test sets using a 70-30 split. It is important to note that data splitting must be completed before any other preprocessing step to ensure no data leakage and that the integrity of the test set, a portion of data that is entirely unseen during model training, is upheld. During splitting, the datasets were stratified to ensure that the same proportion of normal to defective data exists in both. This is an important consideration as a differing distribution of data during the training and inference phases can lead to a performance degradation due to model drift \cite{manias2023model}. The additional preprocessing steps include data standardization (applied consistently across all experiments), and Principal Component Analysis (PCA) used for dimensionality reduction (applied to certain experiments).

Four ML algorithms were considered in this work, Support Vector Machine (SVM), Decision Tree (DT), Random Forest (RF), and $k$-Nearest Neighbors (KNN). These algorithms were selected as a representative sample of ML methods since they include linear, tree-based, ensemble-based, and neighbourhood-based algorithms. A description of the experiments conducted and an analysis of the results obtained for each are presented in the following section.  

\section{Results and Analysis}

The following section will outline the results obtained during this work and provide a comprehensive analysis of the various factors affecting the performance of the ML models. The three sets of results presented include a PCA analysis, a feature isolation analysis, and a feature importance analysis.

\subsection{PCA Analysis}
\begin{figure*}[!hbtp]

\begin{subfigure} {0.61\columnwidth}
    \centerline{\includegraphics[width=\columnwidth]{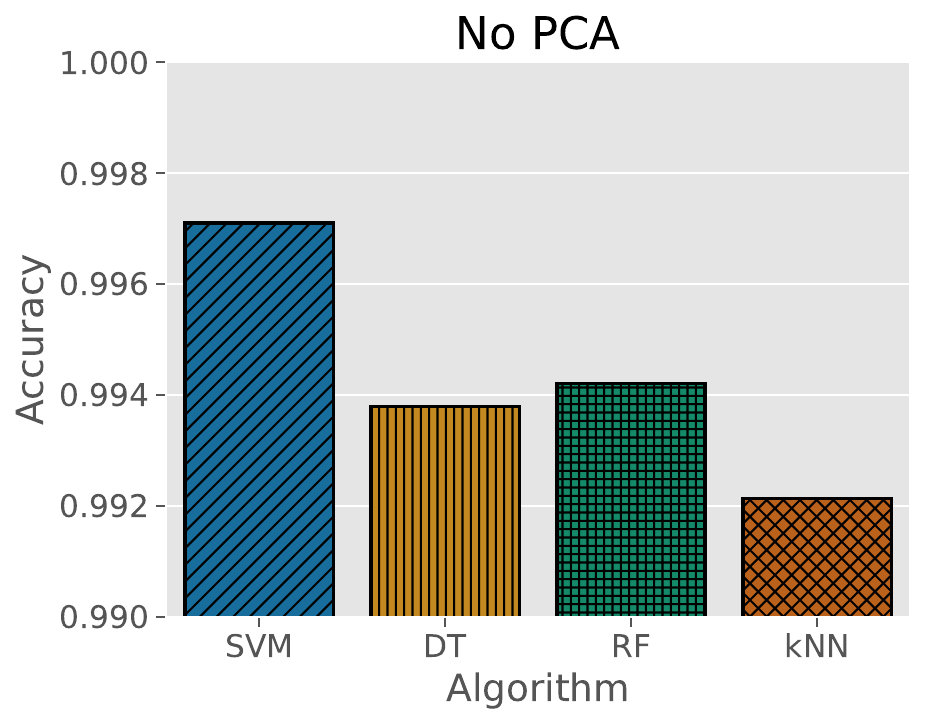}}
    \caption{Algorithm accuracy without PCA.}
    \label{no_pca}
\end{subfigure}
\hfill
\begin{subfigure} {0.61\columnwidth}
    \centerline{\includegraphics[width=\columnwidth]{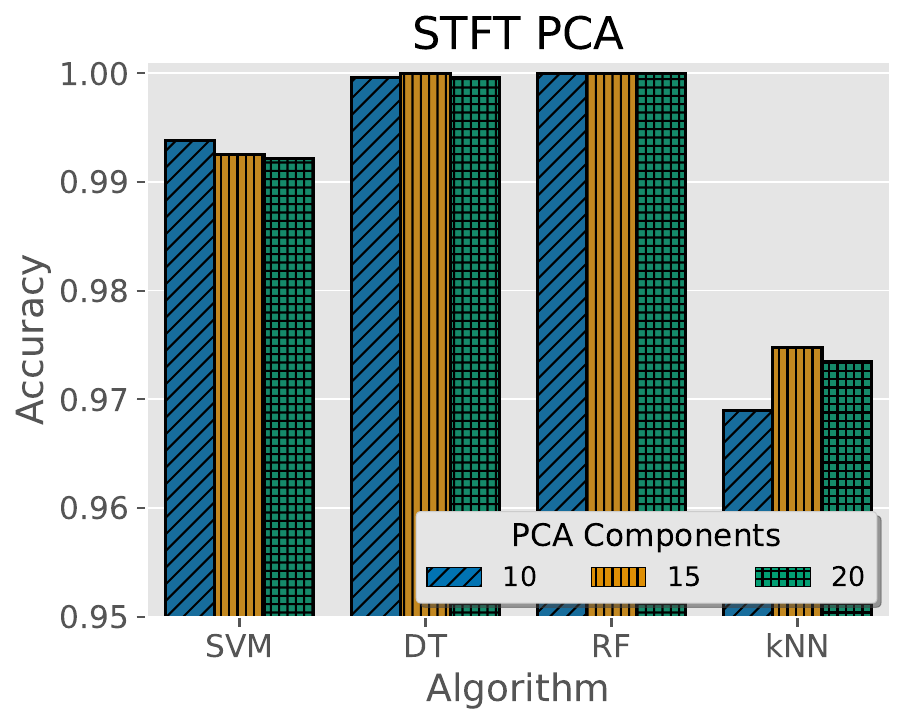}}
    \caption{Algorithm accuracy with STFT PCA.}
    \label{stft_pca}
\end{subfigure}
\hfill
\begin{subfigure} {0.61\columnwidth}
    \centerline{\includegraphics[width=\columnwidth]{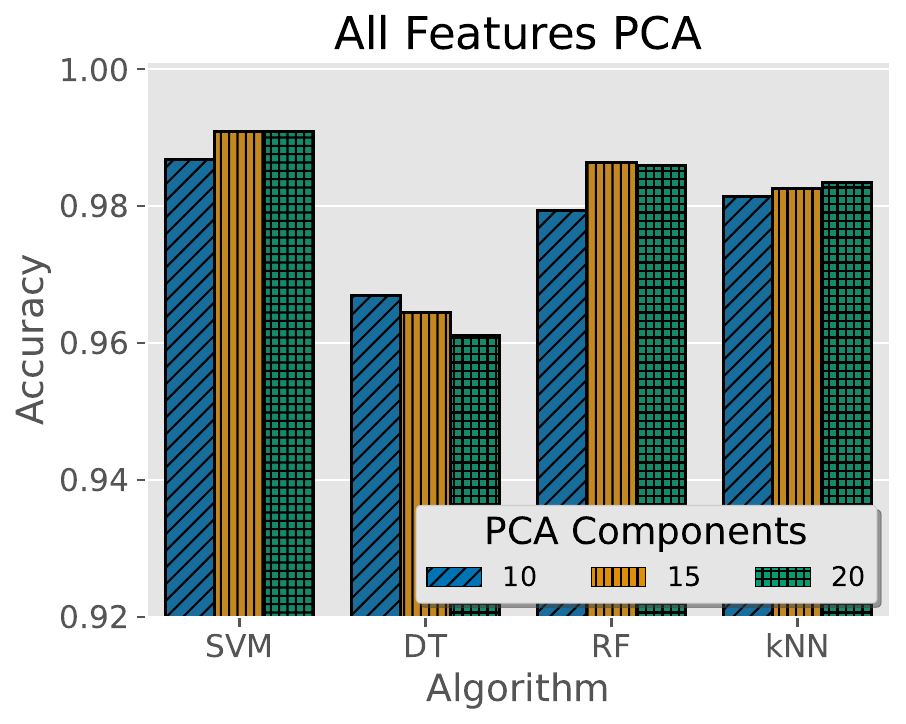}}
    \caption{Algorithm accuracy with all PCA.}
    \label{all_pca}
\end{subfigure}
\caption{PCA Analysis: impact on ML model performance.}
\label{pca_analysis}
\end{figure*}

The first set of results presented in Fig. \ref{pca_analysis} consider a PCA analysis on the feature set, specifically the effect of applying PCA to the feature set on the ML model performance. For this analysis, three experiments were conducted; No PCA applied, PCA applied to the STFT features only, and PCA applied to all features. It should be noted that all features are standardized in all experiments to prevent magnitude biases. The reasoning behind selecting these three experiments is as follows:
\begin{itemize}
\item No PCA: represents a do-nothing scenario where the entirety of the feature set is used.
\item STFT PCA: targets the reduction of the largest subset of features comprising the majority of the feature set.
\item All PCA: explores the reduction of the entire feature set.
\end{itemize}

The results of the No PCA experiment are presented in Fig. \ref{no_pca}. As seen in this figure, all models exhibit performance greater than 99\% accuracy, with the SVM algorithm performing the best. The reasoning behind using accuracy as a metric for comparison is to get a general understanding of the model’s performance. Despite having a minor class imbalance (70 defective to 30 normal), accuracy is still a valid metric for the analysis since the baseline performance of this model is the case where the dominant class (in this case, the defective class) is predicted all the time, resulting in 70\% accuracy. Since the accuracies obtained are significantly higher, it is determined that accuracy is a suitable metric for this analysis. Future work will explore the use of additional metrics (\textit{e.g.,} balanced accuracy, AUC, \textit{etc.}) when considering scenarios where the distribution of labels is severely unbalanced.\par
The results of the STFT PCA experiment are presented in Fig. \ref{stft_pca}. This experiment reduces the vast feature set attributed to the STFT function to 10, 15, and 20 principal components. As seen in Fig. \ref{stft_pca} this has mixed results across the algorithms. In the case of the tree-based algorithms, this results in a performance improvement with accuracy approaching or reaching 100\%. It should be noted that in the case of the decision tree, when 15 principal components are selected, 100\% accuracy is attained. In contrast, when 10 or 20 principal components are selected, the accuracy was approaching but did not reach 100\%. This led us to conduct a PCA analysis, where the number of components was varied between 2 and 30 and the effect of the number of components was assessed. For brevity, these results are not visually presented in this work; however, 15 principal components was determined to be the optimal number for the decision tree. Regarding the random forest algorithm, the use of 10, 15, or 20 principal components resulted in 100\% accuracy. Intuitively, the combination of PCA and tree-based algorithms is expected to increase performance on such a massive feature set since PCA works to maximize the variance while severely reducing the number of features. In the case of the SVM and the kNN algorithms, the use of PCA on the STFT features has negatively affected the performance of the models, suggesting that some useful information specifically affecting the decision boundaries of these models was lost during dimensionality reduction. \par
Finally, the results related to the application of PCA on the entirety of the feature set are shown in Fig. \ref{all_pca}. In these results, we can see that the application of PCA across all features negatively affects all algorithms, suggesting that critical information has been lost. This observation inspired the next set of results, which consider isolating key feature subsets and assessing their individual effect on model performance. \par

\subsection{Feature Isolation Analysis}
\begin{figure*}[!hbtp]

\begin{subfigure} {0.61\columnwidth}
    \centerline{\includegraphics[width=\columnwidth]{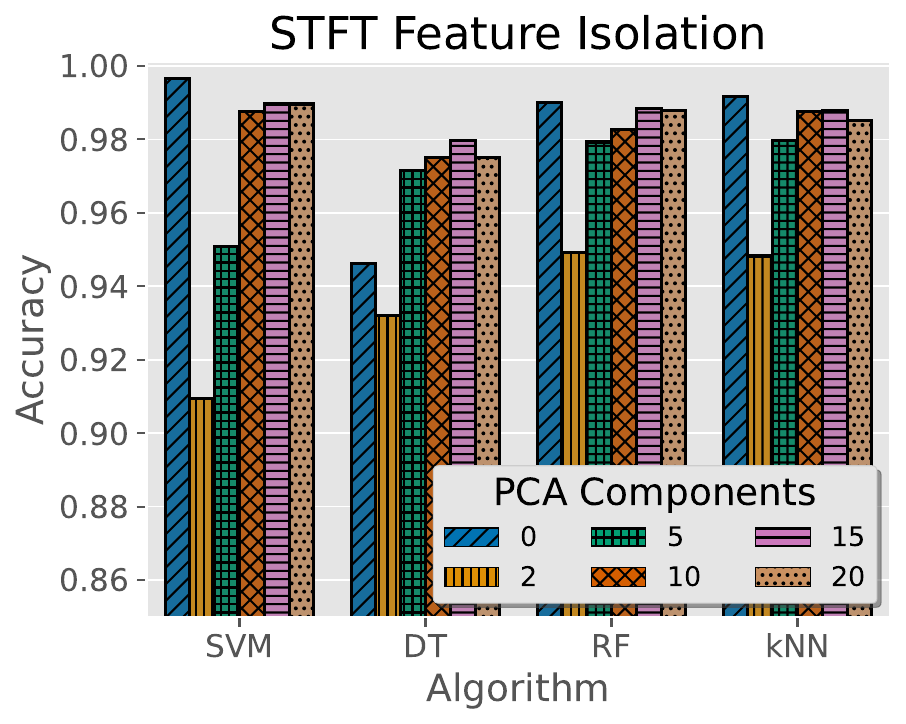}}
    \caption{STFT feature isolation.}
    \label{stft_iso}
\end{subfigure}
\hfill
\begin{subfigure} {0.61\columnwidth}
    \centerline{\includegraphics[width=\columnwidth]{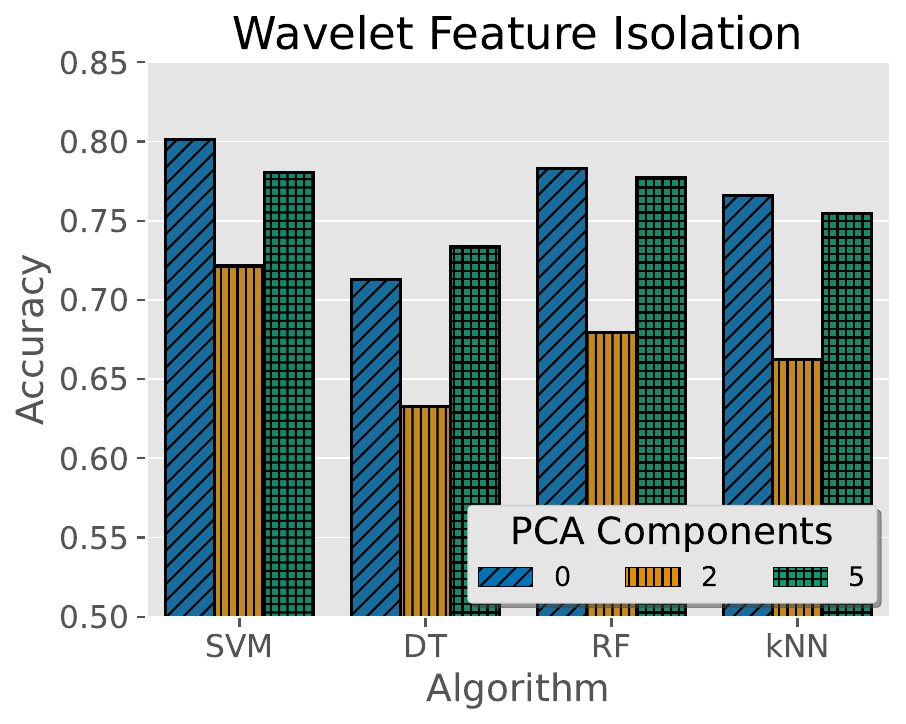}}
    \caption{Wavelet feature isolation.}
    \label{wave_iso}
\end{subfigure}
\hfill
\begin{subfigure} {0.61\columnwidth}
    \centerline{\includegraphics[width=\columnwidth]{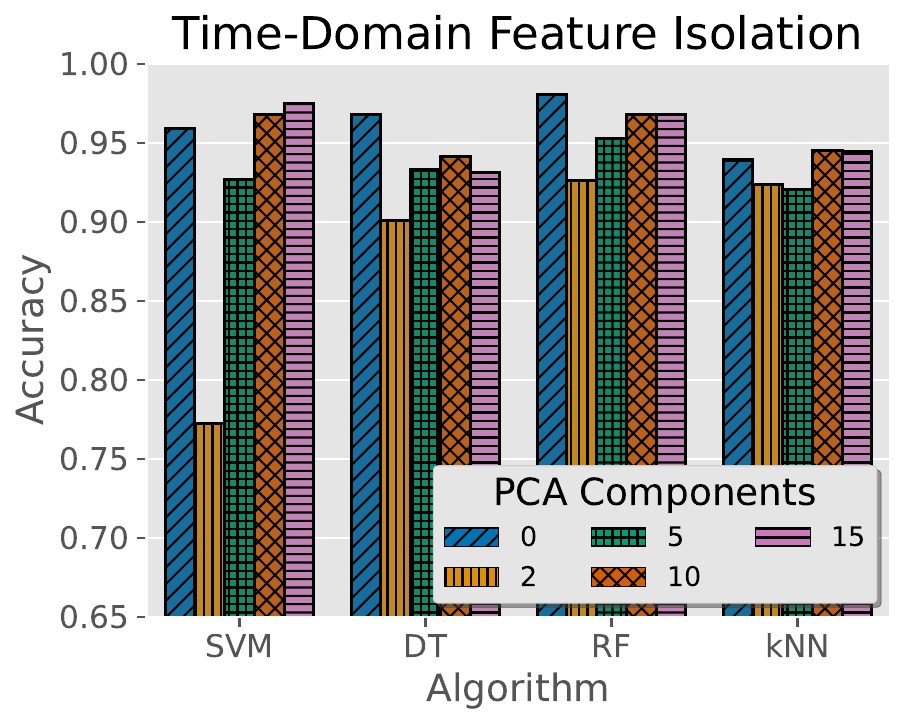}}
    \caption{Time-based feature isolation.}
    \label{time_iso}
\end{subfigure}
\caption{Feature isolation analysis: impact on ML model performance.}
\label{iso_analysis}
\end{figure*}

The second set of results, presented in Fig. \ref{iso_analysis}, consider a feature isolation analysis. In this analysis, three key feature subsets (STFT, Wavelet, and Time-Based) were selected and used to train models where each subset acts as its own feature set. The objective of these experiments is to see how a model trained on a single subset of the feature set performs as an indication of the overall significance of the subset on the classification decision boundary. \par

Figure \ref{stft_iso} presents the STFT feature isolation with PCA applied with a varying number of principal components. As seen, in the case of the SVM, Random Forest, and kNN algorithms, using the entirety of the STFT feature set performs better than the dimensionally reduced STFT feature set. Conversely, when considering the DT algorithm, the use of dimensionality reduction on the STFT feature set resulted an improvement in performance. Only using the STFT features with varying levels of dimensionality reduction resulted in best-performing accuracies exceeding 97\% across all algorithms, signifying the importance of this feature set to the classification decision.

The wavelet feature isolation is presented in Fig. \ref{wave_iso}. It should be noted that since this feature set is already small, the number of principal components during the PCA is restricted. These results show that using only the wavelet features yields a best performance of slightly better than the baseline model performance across all algorithms. This suggests that this feature set is likely not a major contributor to the overall classification decision boundary.

Finally, Fig. \ref{time_iso} considers the isolation of the feature set consisting of all the time-based features. In these results, it can be seen that models built only using the time-based feature set exhibit a best performance that is better than the wavelet feature set but not as good as the STFT feature set. The time-based feature set, in general, performs significantly better than the baseline model performance, indicating that it could contribute to the overall classification decision boundary.

The results presented in this section provided critical insight into the impact of certain feature sets on the overall classification performance. The final set of results will take this analysis a step further and analyze the feature importances of the DT and RF since they were the overall best performing algorithms. 

\subsection{Feature Importance Analysis}
One of the advantages of tree-based algorithms is their explainability through the ability to rank features based on their contribution to determining the optimal decision boundary, commonly known as the feature importance. Through this ranking, a more granular understanding of the effect of certain features (and types of features) on the overall performance is gained. For the purposes of this analysis, the Gini importance is used. It should be noted that these feature importances are determined using the models that do not have PCA since dimensionality reduction reduces model explainability.\par

\begin{figure}[!hbtp]
    \centerline{\includegraphics[width=0.635\columnwidth]{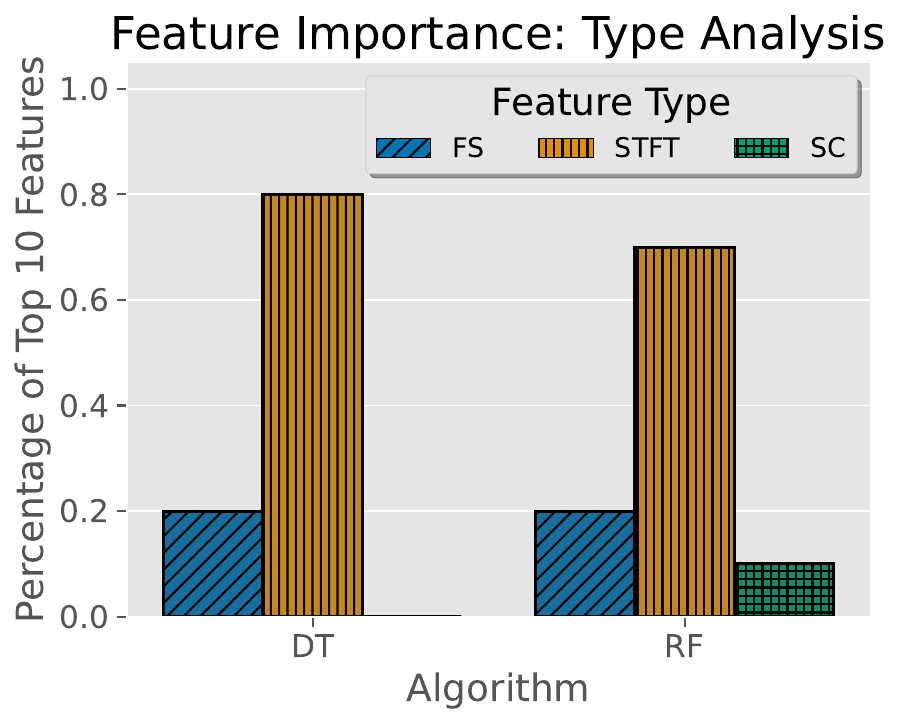}}
    \caption{Top feature type analysis.}
    \label{type_analysis}
\end{figure}

\begin{figure}[!hbtp]
    \centerline{\includegraphics[width=0.635\columnwidth]{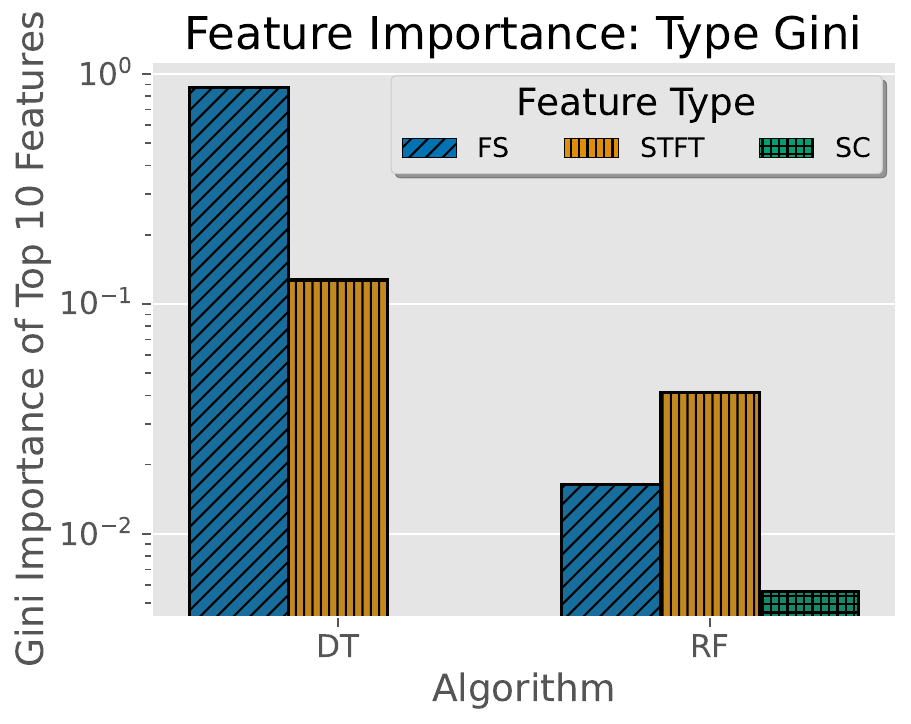}}
    \caption{Top feature type importance analysis.}
    \label{type_importance_analysis}
\end{figure}

The first analysis conducted is the feature type analysis as presented in Fig. \ref{type_analysis} and Fig. \ref{type_importance_analysis}. In this analysis, the type of feature is determined based on the feature subset to which it belongs. In these results, FS refers to Frequency Skewness, and SC refers to Spectral Centroid. As seen in Fig. \ref{type_analysis}, most of the top 10 most important features in the DT and RF algorithms belong to the STFT feature set. Additionally, the top 10 important features in the DT are exclusively limited to the STFT and FS feature sets, whereas the RF also includes a feature from the SC feature set. Figure \ref{type_importance_analysis} explores the magnitude of importance of each identified important feature set. An interesting result is seen regarding the DT algorithm since despite the STFT feature type being the predominant type in the top 10, the FS feature type is the more significant type as its importance, as a type, is almost 100\%. Conversely, in the case of the RF, the feature importance is more analogous to the distribution of feature types in the top 10. \par

\begin{figure}[!hbtp]
     \centerline{\includegraphics[width=0.635\columnwidth]{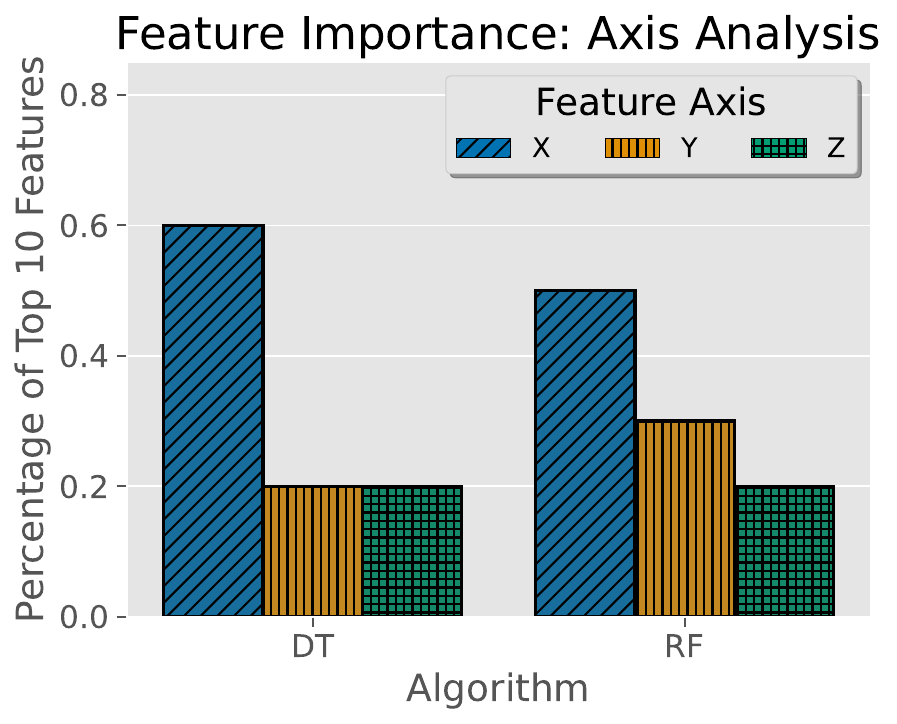}}
    \caption{Top feature axis analysis.}
    \label{axis_analysis}
\end{figure}

\begin{figure}[!hbtp]
\centerline{\includegraphics[width=0.635\columnwidth]{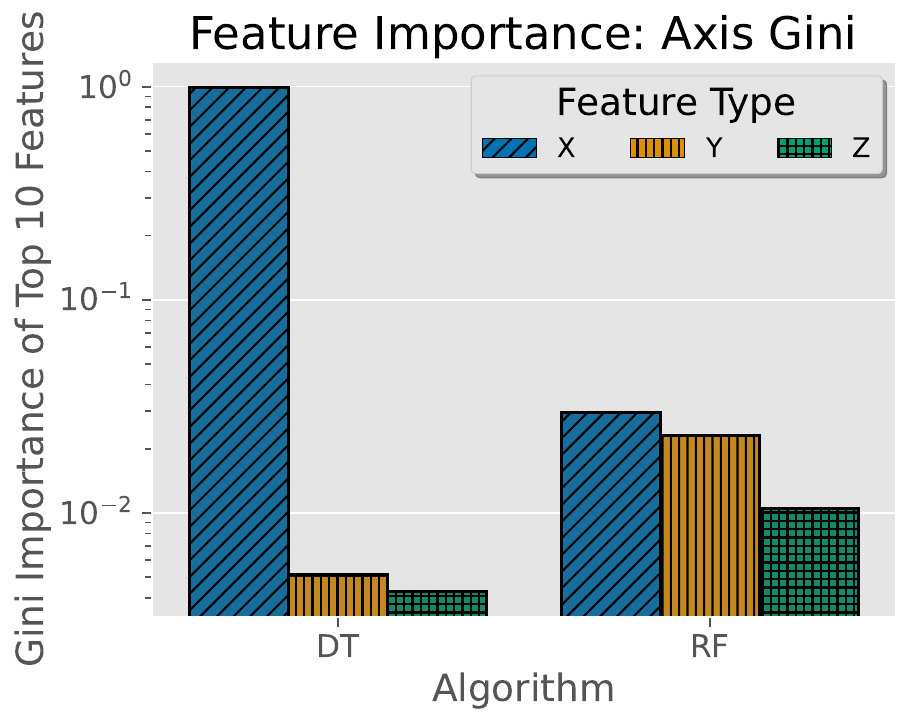}}
    \caption{Top feature axis importance analysis.}
    \label{axis_importance_analysis}
\end{figure}






The second analysis explores the feature importance in terms of the axis it considers, as seen in Fig. \ref{axis_analysis} and Fig. \ref{axis_importance_analysis}. The objective of this analysis is to determine if there is an axis that is more important in distinguishing between normal and defective rotor blades. As seen in Fig. \ref{axis_analysis}, both the DT and RF algorithms use features from all three axes; however, the majority of the features consider the x-axis in both cases. When looking at the importance of the feature sets, the DT algorithm shows a major imbalance as the x-axis features are almost exclusively responsible for determining the decision boundary, with an overwhelming majority of the feature importance approaching a maximum value of 1.0. The RF, on the other hand, is once again more representative of the previous results depicting the axis feature distribution and, in general, does not approach collective feature importances that approach the maximum, as seen with the DT. This indicates that features outside the top 10 most important features are critical to the performance of the RF, whereas only features within the top 10 are critical to the performance of the DT.\par

\section{Conclusion}

The work presented in this paper uses signal processing methods to conduct a vibration analysis on UAV rotor blades. A comprehensive set of time and frequency domain features are then extracted and used to train ML models to detect the presence of a defect on the rotor. The results show that Random Forest algorithm with PCA for dimensionality reduction exhibited the best performance and achieved 100\% classification accuracy. Furthermore, each feature type and its observation axis were analyzed to understand its impact on the classification decision. Future work will consider transforming this problem into a multi-class classification task where the type of defect is also identified. Moreover, data from additional types of defects, environments with ambient noise, and various hovering/flying conditions will be considered.

\bibliographystyle{IEEEtran}
\bibliography{sample}

\end{document}